\documentclass[final,3p,times]{elsarticle}
\usepackage{graphicx}
\usepackage{amssymb}
\usepackage{amsmath}
\usepackage{hyperref}
\hypersetup{
hypertexnames = false,
colorlinks = true,
urlcolor = blue,
linkcolor = blue,
citecolor = blue
}

\makeatletter
\def\ps@pprintTitle{%
 \let\@oddhead\@empty
 \let\@evenhead\@empty
 \def\@oddfoot{}%
 \let\@evenfoot\@oddfoot}
\makeatother

\begin{document}

\begin{frontmatter}

\title{OpenMP solver for rotating spin-1 spin-orbit- and Rabi-coupled\\ Bose-Einstein condensates}

\author[bdu]{Paulsamy Muruganandam}
\ead{anand@bdu.ac.in}

\author[scl]{Antun Bala\v{z}}
\ead{antun@ipb.ac.rs}

\author[ift]{Sadhan K. Adhikari\corref{author}}
\ead{sk.adhikari@unesp.br}
\cortext[author]{Corresponding author.}

\address[bdu]{Department of Physics, Bharathidasan University, Palkalaiperur Campus, Tiruchirappalli 620024, Tamilnadu, India}

\address[scl]{Institute of Physics Belgrade, University of Belgrade, Pregrevica 118, 11080 Belgrade, Serbia}
 
\address[ift]{Instituto de F\'{\i}sica Te\'{o}rica, UNESP -- Universidade Estadual Paulista, 01.140-70 S\~{a}o Paulo, S\~{a}o Paulo, Brazil}

\begin{abstract}
We present OpenMP version of a Fortran program for solving the Gross-Pitaevskii equation for a harmonically trapped three-component rotating spin-1 spinor Bose-Einstein condensate (BEC) in two spatial dimensions with or without spin-orbit (SO) and Rabi couplings. The program uses either Rashba or Dresselhaus SO coupling. We use the split-step Crank-Nicolson discretization scheme for imaginary- and real-time propagation to calculate stationary states and BEC dynamics, respectively.  
\end{abstract}


\end{frontmatter}

\begin{small}
\noindent
{\bf New version program summary}

\noindent\vspace*{-2mm}\\
{\em Program title:} BEC-GP-SPINOR-ROT-OMP, a program package containing programs spin-SO-rot-imre2d-omp.f90, with util.f90.

\noindent\vspace*{-2mm}\\
{\em CPC Library link to program files:} \url{https://doi.org/10.17632/j3wr4wn946.2}

\noindent\vspace*{-2mm}\\
{\em Licensing provisions:} Apache License 2.0

\noindent\vspace*{-2mm}\\
{\em Programming language:} OpenMP Fortran 90/95. The program is tested with the GNU, Intel, PGI, and Oracle (former Sun) compilers.

\noindent\vspace*{-2mm}\\
{\em Supplementary material}: File Supp.pdf gives additional details about the new program version and the underlying physical system.

\noindent\vspace*{-2mm}\\
{\em Journal Reference of previous version}: \href{https://doi.org/10.1016/j.cpc.2020.107657}{Comput. Phys. Commun. {\bf 259} (2021) 107657.}

\noindent\vspace*{-2mm}\\
{\em Does the new version supersede the previous version?}: Only partially. The program spin-SO-rot-imre2d-omp.f90 supersedes spin-SO-imre2d-omp.f90, while the one-dimensional program is not part of this package.

\noindent\vspace*{-2mm}\\
{\em Nature of problem:}
The present Open Multi-Processing (OpenMP) Fortran program solves the time-dependent nonlinear partial differential Gross-Pitaevskii (GP) equation for a trapped rotating spinor Bose-Einstein condensate (BEC) in two spatial dimensions.

\noindent\vspace*{-2mm}\\
{\em Solution method:}
We employ the split-step Crank-Nicolson scheme to discretize the time-dependent GP equation in space and time. The discretized equation is then solved by imaginary- or real-time propagation, employing adequately small space and time steps, to yield the solution of stationary and non-stationary problems, respectively.

\noindent\vspace*{-2mm}\\
{\em Reason for new version:}
The BEC is a special form of matter called superfluid. A hallmark of superfluidity is the formation of quantized vortices in a rotating BEC. The present program can be used to study the generation of quantized vortices in a rotating spin-1 trapped BEC and hence should be of general interest to researchers from various fields.

\noindent\vspace*{-2mm}\\
{\em Summary of revisions:}
Previously we published Fortran \cite{bec2009} and C \cite{bec2012} programs for solving the mean-field GP equation for a BEC, which  are now enjoying widespread use. Later we extended these programs to the more complex scenario of dipolar BECs \cite{dbec2015}, spin-1 spinor BECs \cite{spinor}, and of rotating BECs \cite{vor-lat}. The OpenMP \cite{bec2016,bec2017x} and CUDA/MPI \cite{dbec2016,dbec2016a,dbec2016b} versions of these programs, designed to make these faster and more efficient in multi-core computers, are also available. 

\end{small}

\newpage

In this paper we present Fortran 90/95 program for solving the GP equation of a two-dimensional (2D) rotating spin-1 spinor BEC with Rashba \cite{SOras} and Dresselhaus \cite{SOdre} spin-orbit (SO) coupling and Rabi coupling, involving a modification over the same for a spin-1 spinor BEC \cite{spinor}. A new input parameter OMEG, which represents the angular velocity of rotation $\Omega$ of the spin-1 spinor BEC, has been introduced in the program, following Ref.~\cite{vor-lat}. Besides this new parameter, the execution of the present program follows the same procedure as the 2D program of Ref.~\cite{spinor}. All other input parameters in the two programs are identical and the reader is advised to consult that reference for further details. 

For some values of input parameters the quantized vortices of a rotating BEC could be arranged in the form of a lattice with a certain spatial symmetry, e.g., triangular or square lattice \cite{vor-lat}. In our numerical study, we established recently such a symmetric lattice structure for a Rashba SO-coupled rotating spin-1 BEC in the simplest case, without the Rabi coupling \cite{jpcm}. A Dresselhaus SO-coupled rotating spin-1 BEC should also lead to identical structure, provided the sign of the angular velocity of rotation is changed. For the sake of completeness, in the supplementary material related to this article that can be found online at URL we provide the corresponding GP equations for a rotating spin-1 BEC with some instructive numerical examples.

The program package BEC-GP-SPINOR-ROT-OMP contains programs spin-SO-rot-imre2d-omp.f90 and util.f90 in the directory src, as well as the files makefile and README.md. The makefile allows automated compilation of the program using different supported compilers (GNU, Intel, PGI, Oracle) by a simple make command, as in Ref.~\cite{spinor}. The file README.md contains instructions on how to compile and run the programs. The directory output contains examples of matching outputs of imaginary- and real-time propagation programs in sub-directories with a generic name rot$x$gam$y$ferro or rot$x$gam$y$antiferro, 
where $x$ denotes the value of the angular velocity of rotation $\Omega$ and $y$ denotes the strength of the SO coupling $\gamma$
for ferromagnetic ($c_0 = 482, c_2 = 15$) and antiferromagnetic ($c_0=669, c_2=-3.1 $) cases. The results in imaginary-time sub-directories rot.3gam.5ferro and rot.3gam.5antiferro are calculated using the respective converged imaginary-time wave functions with zero angular velocity. 
The real-time sub-directories rot.3gam.5ferro and rot.3gam.5polar contain real-time results calculated using the respective converged imaginary-time wave functions as inputs.

\section*{Acknowledgments}
\noindent
The work of P.M. forms parts of sponsored research projects by Council of Scientific and Industrial Research (CSIR), India under Grant No. 03(1422)/18/EMR-II, and Science and Engineering Research Board (SERB), India under Grant No. CRG/2019/004059.
A.B. acknowledges funding provided by the Institute of Physics Belgrade, through the grant by the Ministry of Education, Science, and Technological Development of the Republic of Serbia. 
S.K.A. acknowledges support by the CNPq (Brazil) grant 301324/2019-0, and by the ICTP-SAIFR-FAPESP (Brazil) grant 2016/01343-7.

\newpage

\setcounter{equation}{0}
\renewcommand{\theequation}{S.\arabic{equation}}

\setcounter{figure}{0}
\renewcommand{\thefigure}{S\arabic{figure}}

\begin{center}
{\bf\large Supplementary Material}
\end{center}

For tight harmonic binding along the $z$ direction, assuming
a Gaussian density distribution in the $z$ direction, after
integrating out the $z$ coordinate, in the mean-field approximation, a quasi-2D rotating SO-coupled spin-1 spinor BEC is described by the
following set of three coupled GP equations for $N$ atoms, of mass $\widetilde m$ each,
in a dimensionless form for the hyper-fine spin components
 $F_z = \pm 1, 0$ \cite{thspinor,thspinorb,GA}
\begin{align}\label{EQ1} 
i {\partial_t \psi_{\pm 1}({\bf r})}&= \left[{\cal H}+{c_2}
\left(\rho_{\pm 1} -\rho_{\mp 1} +\rho_0\right) \right] \psi_{\pm 1}({\bf r})+\left\{c_2 \psi_0^2({\bf r})\psi_{\mp 1}^*({\bf r})\right\} + \frac{\cal R}{\sqrt 2}\psi_{0}({\bf r})+\gamma f_{\pm 1}({\bf r}) \, , 
\\ \label{EQ2}
i {\partial_t \psi_{0}({\bf r})}&=\left[{\cal H}+{c_2}
\left(\rho_{+ 1}+\rho_{- 1}\right)\right] \psi_{0}({\bf r})
+\left \{ 2c_2 \psi_{+1}({\bf r})\psi_{-1}({\bf r})\psi_{0}^* ({\bf r})\right\}+ \frac{\cal R}{\sqrt 2} \sum_{j=+1,-1} \psi_{j}({\bf r})
+\gamma f_0({\bf r})  \, , \\
{\cal H}&= -\frac{1}{2}\nabla_{\bf r}^2+V({\bf r})+c_0 \rho-\Omega L_z\, , \\
c_0&=\frac{2N\sqrt{2 \pi \kappa}(a_0+2a_2)}{3l_0}, \quad c_2=\frac{2N\sqrt{2 \pi \kappa}(a_2-a_0)}{3l_0}
\end{align}
where $\rho_j = |\psi_j|^2$ are the densities of components $j= \pm 1 , 0$, and $\rho ({\bf r})= \sum \rho_j({\bf r})$ is the total density, $V({\bf r})$ is the confining trap,  
$\partial_t$  is the partial derivative with respect to time,  $\Omega $ is the velocity of rotation in units of the harmonic trap frequency $\omega$ in the $x-y$ plane, $l_0=\sqrt{\hbar/\widetilde m \omega}$,
$L_z$ is the $z$ component of angular momentum, $ {\bf r}= \{ x,y \}, \nabla_{\bf r}^2 = \partial_x^2 +\partial_y^2$
and  ${\cal R}$ ($\gamma$) is the strength of Rabi (SO) coupling, with $a_0$ and $a_2$ being the scattering lengths in the total spin channels 0 and 2.  Here $\kappa = \omega_z/\omega$ $(\gg 1)$,  
where $\omega_z$ is the harmonic trap frequency in the $z$ direction.
In Eqs.~(\ref{EQ1}) and (\ref{EQ2}), lengths are expressed in units of $l_0$, condensate density in units of $l_0^{-2}$, and time in units of $\omega^{-1}$. 
We consider the general SO coupling term in the form $\gamma f\equiv \gamma(\eta p_y \Sigma_x - p_x \Sigma_y)$, where $\eta=1,-1$  for Rashba \cite{SOras}, Dresselhaus \cite{SOdre} SO couplings, and where $p_x = -i \partial_x, p_y=-i\partial_y$ are the momentum operators. $\Sigma_x$,
$\Sigma_y$  are the irreducible representations of the $x, y$  components of the spin-1 matrix $\Sigma$, with components
\begin{align}\label{smat}
\Sigma_x= \frac{1}{\sqrt 2} \left( \begin{array}
 {ccccc}
0 & 1 & 0\\
1 & 0 & 1\\
0 & 1 & 0 \end{array} \right)\, , \quad 
\Sigma_y= \frac{i}{\sqrt 2} \left( \begin{array}
 {ccccc}
0 & -1 & 0\\
1 & 0 & -1\\
0 & 1 & 0 \end{array} \right)\, .
\end{align}
In Eqs.~(\ref{EQ1}) and (\ref{EQ2}), the Rashba and Dresselhaus coupling terms in 2D are $\gamma  f_{\pm 1}({\bf r})= - {i\widetilde \gamma}\Big[\eta \partial_y \psi_0 ({\bf r}) \pm i \partial_x \psi_0 ({\bf r}) \Big]$ and 
$\gamma f_0({\bf r}) = - {i\widetilde \gamma} \Big[-i \partial_x \psi_{+1}({\bf r}) 
+i\partial_x \psi_{-1}({\bf r}) + \eta\partial_y \psi_{+1}({\bf r}) +
\eta \partial_y \psi_{-1}({\bf r}) 
\Big] $, with $\widetilde \gamma = \gamma /\sqrt 2$.  
 
The normalization and magnetization ($m$) conditions are given by
$ \int \rho({\bf r})\, d{\bf r}=1$ and $\int \Big[\rho_{+1}({\bf r})
-\rho_{-1}({\bf r})\Big] \, d{\bf r}=m$.
The energy functional of the system is \cite{thspinor,thspinorb}
\begin{align}\label{energy}
E=&\frac{1}{2} \int d {\bf r}\, \Bigg\{\sum_j |\nabla _{\bf r}\psi_j|^2+2V\rho + c_0 \rho^2-2\Omega \sum_j
\psi_j^*L_z\psi_j
\nonumber \\ &
+ c_2\left[ \rho_{+1}^2+ \rho_{-1}^2 +2 \left( \rho_{+1}\rho_0+ \rho_{-1}\rho_0- \rho_{+1}\rho_{-1}+\psi_{-1}^*\psi_0^2\psi_{+1}^*+
\psi_{-1}{\psi_0^*}^2\psi_{+1}
\right) 
\right] \nonumber \\ &
+ {{\sqrt 2}\cal R}\Big[  (\psi^*_{+1}+ \psi^*_{-1})\psi _0+ \psi_0^* (\psi_{+1}+ \psi_{-1})    \Big]   
+ {2}\gamma {\Big[}\psi_{+1}^*f_{+1}+ \psi_{-1}^*f_{-1} +
 \psi_0^* f_0   {\Big]}   \Bigg\}.
\end{align}

All calculations reported here were performed with the predefined  space and time steps DX and DT in the programs: 
DX = DY = 0.1, DT = DX*DX*0.1 (imaginary time) and DT = DX*DX*0.025 (real time). To increase the accuracy of calculation, 
the space steps DX and DY should be reduced and the total number of space discretization points NX and NY increased proportionally.

The parameters of the GP equation $c_0$ and $c_2$ are taken from the following realistic experimental situations.  
For the ferromagnetic BEC we use the following parameters of $^{87}$Rb atoms: 
$N=10^5, l_0/\sqrt \kappa = 4$ $\mu$m, $a_0=101.8 a_B, a_2=100.4 a_B$ \cite{kok}, where $a_B$ 
is the Bohr radius. Consequently, $c_0\equiv 2N\sqrt{2\pi \kappa }(a_0+2a_2)/3l_0 \approx 669$ and $c_2\equiv  2N\sqrt{2\pi \kappa }(a_2-a_0)/3l_0\approx -3.1$.  
For the quasi-2D anti-ferromagnetic 
BEC we use the following parameters of $^{23}$Na atoms: $N=136,200, l_0/\sqrt \kappa = 4$ $\mu$m, $a_0=50.00 a_B, a_2=55.01 a_B$ \cite{naa}. Consequently, $c_0  \approx 482$ and $c_2 \approx 15$.  
 
\begin{figure}[!t]
\centering
\includegraphics[width=.16\linewidth]{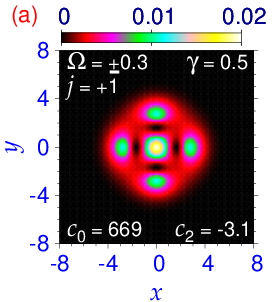}
\includegraphics[width=.16\linewidth]{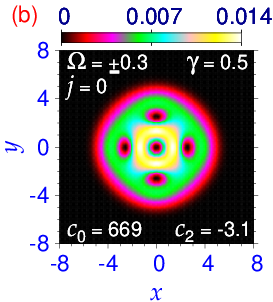}
\includegraphics[width=.16\linewidth]{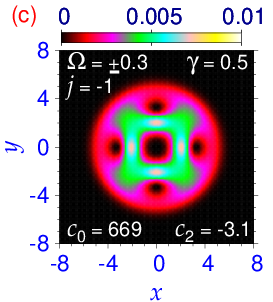}
\includegraphics[width=.16\linewidth]{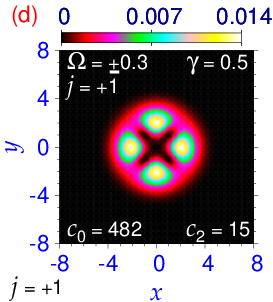}
\includegraphics[width=.16\linewidth]{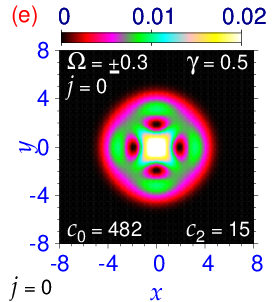}
\includegraphics[width=.16\linewidth]{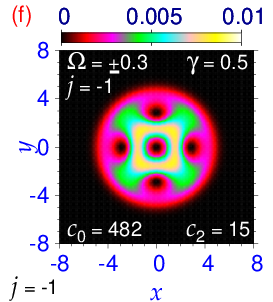}
\includegraphics[width=.16\linewidth]{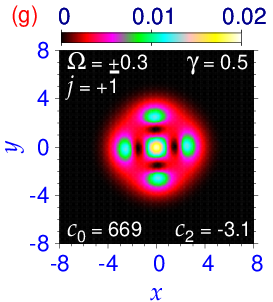}
\includegraphics[width=.16\linewidth]{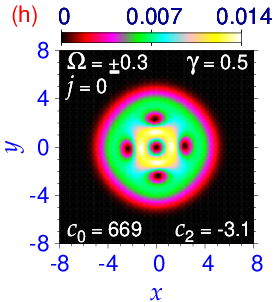}
\includegraphics[width=.16\linewidth]{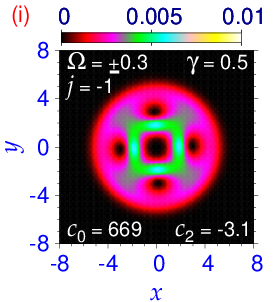}
\includegraphics[width=.16\linewidth]{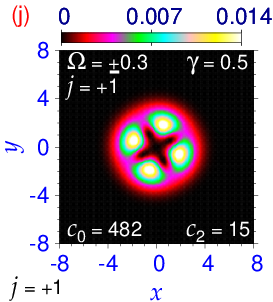}
\includegraphics[width=.16\linewidth]{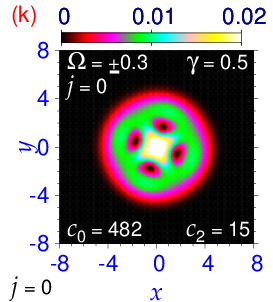}
\includegraphics[width=.16\linewidth]{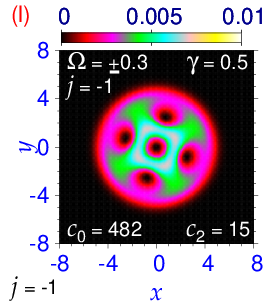}
\caption{Contour plot of component densities  $\rho_j(x, y)$ for (a) $j=+1$, (b) $j=0$, and (c) 
$j=-1$, obtained in imaginary-time simulation of a quasi-2D harmonically trapped Rashba or Dresselhaus SO-coupled ferromagnetic BEC with nonlinearities $c_0=669$, $c_2=-3.1$, angular velocity $\Omega =\pm 0.3$, and SO-coupling strength  $\gamma=0.5$.  
(d)-(f) The same for a Rashba or Dresselhaus SO-coupled anti-ferromagnetic BEC with nonlinearities $c_0=482$, $c_2=15$, and $\Omega=\pm 0.3, \gamma = 0.5$.    (g)-(i)  The same obtained in real-time simulation of the ferromagnetic BEC with nonlinearities $c_0=669$, $c_2=-3.1$, and $\Omega =\pm 0.3, \gamma=0.5$ after 50 units of time. 
 (j)-(l) The same obtained in real-time simulation of the antiferromagnetic BEC with nonlinearities $c_0=482$, $c_2=15$, and $\Omega =\pm 0.3, \gamma=0.5$ after 50 units of time. In all cases, the positive (negative) angular velocity refers to the Rashba (Dresselhaus) coupling.} 
\label{fig1}
\end{figure}

In the presence of SO coupling the spherically-symmetric ground states of a non-rotating BEC has vortices at the center. For an antiferromagnetic BEC they are of the type $(\mp 1,0,\pm 1)$, where the numbers in parentheses denote the angular momentum of vortices at the center of the components, with the upper (lower) sign referring to the Rashba (Dresselhaus) coupling. For a ferromagnetic BEC, they are of the type $(0,\pm 1,\pm 2)$. The imaginary-time calculation of the SO-coupled non-rotating BEC has these states built-in in the initial wave function. For a rotating SO-coupled BEC, the Rashba and Dresselhaus coupling lead to identical density structure, provided that the sign of a rotation velocity is changed in the case of Dresselhaus coupling. 

In Figs.~\ref{fig1}(a)-(c) we show the contour plot of the density of components $j=+1,0,-1$ obtained by imaginary-time calculation of 
a ferromagnetic BEC with parameters $\Omega=0.3$ and $\gamma =0.5$. The same for the antiferromagnetic BEC is displayed in Figs.~\ref{fig1}(d)-(f). The imaginary-time simulation for $\Omega =0.3$ was done using the converged wave function of the imaginary-time simulation for $\Omega =0$ as the initial state. For small $\gamma$ and $\Omega$ these states are dynamically stable, as shown through the contour plot of the density of respective real-time simulation in Figs.~\ref{fig1}(g)-(i) and Figs.~\ref{fig1}(j)-(l) for the ferromagnetic and anti-ferromagnetic BECs after 50 units of time. The real-time simulation was performed using the respective converged wave function of the imaginary-time simulation as the initial state. The results of real-time simulations oscillate a little, but the vortices remain intact. 

\end{document}